\documentclass[11pt,a4paper]{article}
\usepackage{graphicx}
\usepackage{jcappub,natbib}
\usepackage{braket}

\def\la{~\mbox{\raisebox{-.6ex}{$\stackrel{<}{\sim}$}}~}
\def\ga{~\mbox{\raisebox{-.6ex}{$\stackrel{>}{\sim}$}}~}

\title{
Parity-violating CMB correlators with non-decaying statistical anisotropy}
\author[a,b]{Nicola Bartolo,}
\author[a,b,c]{Sabino Matarrese,}
\author[d]{Marco Peloso}
\author[a,b,e]{and Maresuke Shiraishi}

\affiliation[a]{Dipartimento di Fisica e Astronomia ``G. Galilei'', 
Universit\`a degli Studi di Padova, 
via Marzolo 8, I-35131, Padova, Italy}
\affiliation[b]{INFN, Sezione di Padova, 
via Marzolo 8, I-35131, Padova, Italy}
\affiliation[c]{Gran Sasso Science Institute, INFN, viale F. Crispi 7, I-67100, L'Aquila, Italy}
\affiliation[d]{School of Physics and Astronomy, University of Minnesota, Minneapolis 55455, USA}
\affiliation[e]{Kavli Institute for the Physics and Mathematics of the Universe (Kavli IPMU, WPI), UTIAS, The University of Tokyo, Chiba, 277-8583, Japan}

\abstract{
We examine the effect induced on cosmological correlators by the simultaneous breaking of parity and of statistical isotropy. As an example of this, we compute the scalar-scalar, scalar-tensor, tensor-tensor and scalar-scalar-scalar cosmological correlators in presence of the coupling ${\cal L} = f(\phi) ( - \frac{1}{4} F^2 + \frac{\gamma}{4} F \tilde{F} )$ between the inflaton $\phi$ and a vector field with vacuum expectation value ${\bf A}$. For a suitably chosen function $f$, the energy in the vector field $\rho_{\rm A}$ does not decay  during inflation. This results in nearly scale-invariant signatures of broken statistical isotropy and parity.  Specifically, we find that the scalar-scalar correlator of primordial curvature perturbations includes a quadrupolar anisotropy, $P_\zeta ( {\bf k}) = P(k)[ 1 + g_* ( \hat{\bf k} \cdot \hat{\bf A})^2]$, and a (angle-averaged) scalar bispectrum that is a linear combination of the first $3$ Legendre polynomials, $B_\zeta(k_1, k_2, k_3) = \sum_L c_L P_L (\hat{\bf k}_1 \cdot \hat{\bf k}_2) P(k_1) P(k_2) + 2~{\rm perms} $, with $c_0 : c_1 : c_2 = 2 : -3 : 1$ ($c_1 \neq 0$ is a consequence of parity violation, corresponding to the constant $\gamma \neq 0$). The latter is one of the main results of this paper, which provides for the first time a clear example of an inflationary model where a non-negligible $c_1$ contribution to the bispectrum is generated.  The scalar-tensor and tensor-tensor correlators induce characteristic signatures in the Cosmic Microwave Background temperature anisotropies (T) and polarization (E/B modes); namely, non-diagonal contributions to  $\langle a_{\ell_1 m_1}  a_{\ell_2 m_2}^* \rangle$, with $|\ell_1 - \ell_2| = 1$ in TT, TE, EE and BB, and $|\ell_1 - \ell_2| = 2$ in TB and EB. The latest CMB bounds on the scalar observables ($g_*$, $c_0$, $c_1$ and $c_2$), translate into the upper limit $\rho_{\rm A} / \rho_\phi \lesssim 10^{-9}$ at $\gamma=0$. We find that the upper limit on the vector energy density becomes much more stringent as $\gamma$ grows.  
}

\begin{document}

\begin{flushright}
{\small UMN-TH-3435/15 \\ IPMU15-0067}
\end{flushright}

\maketitle
\flushbottom

\section{Introduction}

Pseudo-scalar fields, which can naturally emerge in global symmetry breakings, have often been employed in models of cosmological inflation  (e.g., see refs.~\cite{Freese:1990rb,Adams:1992bn,Lue:1998mq,Kim:2004rp,Dimopoulos:2005ac,McAllister:2008hb,Kaloper:2008fb,Kaloper:2011jz, Pajer:2013fsa, Dimastrogiovanni:2012st, Dimastrogiovanni:2012ew}). A typical model of a pseudo-scalar coupled to a gauge field is 
\begin{eqnarray}
{\cal L} = 
- \frac{1}{2} \left( \partial \phi \right)^2 - V \left( \phi \right) - \frac{1}{4} F^2  -  \frac{\phi}{4 f} F \tilde{F}  ~, 
\label{eq:Lag_pseudo_normal}
\end{eqnarray}
where $\phi$ is the pseudoscalar field and $1 / f$ expresses the strength of the axial coupling. Due to the motion of the inflaton $\phi (t)$, this coupling  enhances one of the two helicity states of the gauge field during inflation,  inducing parity violation. This violation can be imprinted on  the gravitational waves through the gravitational interactions of the gauge field.  Most common observables of such chiral tensor perturbations are the cross correlations between temperature anisotropies T/E-mode polarization and B-mode polarization (TB/EB) of the Cosmic Microwave Background (CMB) \cite{Lue:1998mq,Sorbo:2011rz, Barnaby:2011vw, Maleknejad:2014wsa, Bielefeld:2014nza}. Since the inflaton typically speeds up during inflation, the production of gauge quanta  can potentially increase at smaller scales, giving rise to a gravity wave signal observable at terrestrial interferometers  \cite{Cook:2011hg, Barnaby:2012xt,Obata:2014loa}. The chiral nature of this signal can be probed by combining measurements from multiple interferometers \cite{Seto:2007tn,Crowder:2012ik}. In addition, the $f (\phi) F \tilde{F}$ interaction can also generate large primordial non-Gaussianity  \cite{Barnaby:2010vf, Barnaby:2011vw, Meerburg:2012id, Cook:2013xea, Shiraishi:2013kxa, Ferreira:2014zia, Ade:2013ydc, Shiraishi:2014ila, Ade:2015ava}, and primordial black holes \cite{Linde:2012bt}. Finally, this coupling has also been employed in models of inflationary magnetogenesis \cite{Garretson:1992vt, Son:1998my, Field:1998hi, Vachaspati:2001nb,Sigl:2002kt, Durrer:2010mq,  Caprini:2014mja, Cheng:2014kga, Fujita:2015iga}.

In the framework of the pseudoscalar inflation, impacts due to statistical anisotropy have been recently analyzed, by introducing a coherent vacuum expectation value (vev) of the gauge field \cite{Dimopoulos:2012av, Maleknejad:2013npa, Bartolo:2014hwa}. In such a case, the primordial correlators are affected by the simultaneous breaking of parity and statistical isotropy. This generates characteristic $\langle a_{\ell_1 m_1}  a_{\ell_2 m_2}^* \rangle$ correlators of the CMB temperature anisotropies and polarization. Specifically, one finds non-vanishing TT, TE, EE and BB correlators for $|\ell_1 - \ell_2| = 1$, and non-vanishing TB and EB correlators for $|\ell_1 - \ell_2| = 2$~\cite{Bartolo:2014hwa}. 
Mathematically, all the CMB signals satisfying $|\ell_1 - \ell_2| = {\rm odd} \ ({\rm even})$ in TT, TE, EE and BB (TE and EB) are parity-odd. These can be realized only when the primordial correlators include parity-violating information. On the other hand, the primordial correlators can generate nonzero CMB signals satisfying $\ell_1 \neq \ell_2$, only when they break rotational invariance. Hence, the above signals become distinctive indicators of broken parity and rotational invariance in the primordial Universe. 
\footnote{These discussions rely on the assumption that there is no mechanism breaking parity and isotropy at late times.}

In ref.~\cite{Bartolo:2014hwa}, we have presented a proof of the existence of these interesting signals, analyzing the simplest lagrangian \eqref{eq:Lag_pseudo_normal}. In this case, however, the vev of the vector field (which is the quantity breaking statistical isotropy) decays very rapidly during inflation,  $ \rho_{\rm A} \propto {\bf A}^2 \propto a^{-4}$ (where $a$ is the scale factor), so that one needs to assume that a vector vev is present as an ad-hoc initial condition when the CMB modes left the horizon. The rapid decrease of $\rho_{\rm A}$ then implies that only the largest CMB modes can be affected by it, so that the induced non-diagonal correlators are highly red-tilted (they are $\propto k^{-4}$). It was found in ref.~\cite{Bartolo:2014hwa} that  only CMB multipoles with $\ell \lesssim 10$ can be affected at a detectable level. 

As anticipated in ref.~\cite{Bartolo:2014hwa}, a more natural initial condition, and a more interesting signal, can be obtained  if the kinetic term of the vector field is modified as in the Ratra mechanism  \cite{Ratra:1991bn}, so to allow for a nearly constant  $\rho_{\rm A}$. Specifically, if the kinetic term is $- \frac{f(\phi)}{4} F^2$, and if the functional form of $f ( \phi )$ is chosen such that the background evolution satisfies $f ( \phi (t) ) = a^{-4} (t)$, then the vector has a constant electric vev (we are using standard electromagnetic notation for simplicity, although we do not need to assume that $A_\mu$ corresponds to the standard model photon). This time dependence can be achieved through a suitable relation between $f ( \phi )$ and the inflation potential  \cite{Martin:2007ue}.

A model that provides a non-decaying $ \rho_{\rm A}$ and that breaks parity has been recently considered in ref.~\cite{Caprini:2014mja} as a model for primordial  magnetogenesis \cite{Caprini:2014mja}. It is characterized by the lagrangian~\footnote{In the proposal of ref.~\cite{Caprini:2014mja}, the field $\phi$ does not need to be the inflaton field.}
\begin{eqnarray}
{\cal L} = - \frac{1}{2} \left( \partial \phi \right)^2 - V ( \phi ) 
+ I^2 (\phi) \left( - \frac{1}{4} F^2
  +  \frac{\gamma}{4} F \tilde{F} 
\right) ~, \label{eq:Lag_pseudo_new}
\end{eqnarray}
 where $\gamma$ is a constant.~\footnote{The explicit breaking of parity can be avoided  if the $F^2$ and the $F \tilde{F}$ term are proportional to two different fields (as for instance in supergravity) that have vevs that evolve during inflation maintaining a constant ratio \cite{Caprini:2014mja}.} Analogously to the well studied $\gamma =0$ case \cite{Himmetoglu:2009mk, Dulaney:2010sq, Gumrukcuoglu:2010yc, Watanabe:2010fh, Maleknejad:2011jr, Maleknejad:2012fw, Maleknejad:2012as, Bartolo:2012sd, Shiraishi:2013vja, Shiraishi:2013oqa, Naruko:2014bxa}, we show in this work that, for $I(\phi) \propto a^{-2}$, the interaction between $\phi$ and $A_\mu$ leaves  nearly scale-invariant signatures on the primordial cosmological correlators. Differently to the $\gamma = 0$ case, however, due to the $I^2 ( \phi) F {\tilde F}$ interaction, one vector helicity state is produced with a greater abundance that the other one. As mentioned above, this violation of parity can affect the CMB correlators through the gravitational interactions of the vector field. 

In addition, a stronger effect arises from the direct coupling between the vector field and the inflaton. We mentioned that the interaction in eq.~\eqref{eq:Lag_pseudo_new}  is responsible for (i) maintaining a nearly constant energy in the vector field during inflation, and (ii)  enhancing one helicity of the vector field with respect to the other one. These effects are due to the classical evolution of $I ( \phi )$ and therefore to the vev of the inflaton field. However the same interaction term \eqref{eq:Lag_pseudo_new} also couples the vector quanta to the inflaton perturbations. As we will see, in this model 
 $\rho_{\rm A} / \rho_\phi \ll 1$ is required. Therefore one can identify the perturbations of the inflaton as the adiabatic perturbation of the model, with negligible error. As we show below, the couplings $\delta \phi \delta A$ and  $\delta \phi \delta A^2$ encoded in eq.~\eqref{eq:Lag_pseudo_new} modify both the power spectrum and the bispectrum of the primordial scalar perturbations. For the power spectrum of primordial curvature perturbations, one obtains the ACW  \cite{Ackerman:2007nb} quadrupolar term 
\begin{equation}
P_\zeta \left( {\bf k} \right) = P \left( k \right) 
\left[ 1 + g_* ( \hat{\bf k} \cdot \hat{\bf A} )^2 \right]
\end{equation}
with a nearly scale invariant $g_*$ parameter. For the  (angle-averaged) bispectrum, one obtains the first three terms of an expansion in Legendre polynomials  \cite{Shiraishi:2013vja}: 
\begin{equation}
\label{eq:BinLegendre}
B_\zeta(k_1, k_2, k_3) = \sum_{L=0}^2 c_L P_L (\hat{\bf k}_1 \cdot \hat{\bf k}_2) P(k_1) P(k_2) + 2~{\rm perms.}\, , 
\end{equation}
with $c_0 : c_1 : c_2 = 2 : -3 : 1$.~\footnote{Notice that, due to a non-vanishing vev of the vector field, a statistical anisotropic bispectrum is actually generated and, after an angular average (see the discussion after eq.~\eqref{eq:c_L_def}), it takes the form~\eqref{eq:BinLegendre}. For studies of CMB bispectra that break statistical isotropy see refs.~\cite{Bartolo:2011ee,Shiraishi:2011ph}.}  The three parameters $\{g_*, c_0, c_2\}$ arise due to the breaking of statistical isotropy, and were already found in the $- \frac{f(\phi)}{4} F^2$ model~\cite{Bartolo:2012sd,Shiraishi:2013vja}. The non-vanishing of $c_1$ instead requires both breaking of statistical isotropy and parity.  This is the first time that, within these conditions, a non-negligible contribution $c_1$ to the curvature bispectrum is generated \emph{during} inflation.~\footnote{See refs.~\cite{Shiraishi:2012sn, Shiraishi:2013vja} for discussions on another possibility to generate non-negligible $c_1$ from large-scale helical magnetic fields at the radiation dominated era.}

As we show in section~\ref{subsec:observable}, for $\gamma = 0$, the fact that such signatures have not been observed in the CMB data enforces the upper limit $\rho_{\rm A} / \rho_\phi \la 10^{-9}$.  In our work we study the $\gamma \neq 0$ case. Our results are summarized in figure \ref{fig:95CL_gamma_Ephi}, where we show the upper limit on  $\rho_{\rm A} / \rho_\phi$ as a function of $\gamma$, starting from $\gamma > 1$. Strictly speaking, our analytic computation is valid for $\gamma \gg \frac{1}{2}$. However, the limit is continuous in $\gamma$, and figure \ref{fig:95CL_gamma_Ephi} shows that extrapolating our lines in the $\gamma \la 1$ intervals provides results consistent with the limit at $\gamma = 0$. To give a measure on how strongly the upper limit decreases with $\gamma$, we note that our computation provides the upper limit $\rho_{\rm A} / \rho_\phi \la 4 \times 10^{-16}$ at $\gamma =1$, and $\rho_{\rm A} / \rho_\phi \la 3 \times 10^{-25}$ at $\gamma =2$.

Although having such a small $\rho_{\rm A}$ is a mathematical possibility, in section~\ref{subsec:expected} we argue that a greater energy in the background vector field  should be expected, simply from the addition of IR modes \cite{Bartolo:2012sd}. This essentially rules out the model \eqref{eq:Lag_pseudo_new}, under the assumption that it produces a constant vector energy density, that $\gamma > {\rm O } ( 5 )$, and that $\phi$ is the inflaton. Our results do not immediately extend to the case in which $\phi$ is not the inflaton. However, we expect that also in that case the primordial perturbations will be significantly affected by the effects we have studied, mostly due to the linear coupling (which exists at least due to gravity) between $\delta \phi$ and the inflaton perturbations \cite{Ferreira:2014zia}. 

This paper is organized as follows. In the next section, we review an inflationary model based on the lagrangian \eqref{eq:Lag_pseudo_new} and we discuss how it can lead to a constant vev of the gauge field. In section~\ref{sec:correlator}, we compute the scalar-scalar, scalar-tensor, tensor-tensor and scalar-scalar-scalar correlators of primordial curvature perturbations, and find the distinct observable predictions of the lagrangian \eqref{eq:Lag_pseudo_new}. In section~\ref{sec:constraints}, we estimate the observational bounds on the energy density of the gauge field vev from the latest {\it Planck} constraints on $g_*$, $c_0$, $c_1$ and $c_2$. This result is discussed in section~\ref{sec:constraints} and in the concluding section.

\section{A model for breaking parity and statistical isotropy}

Let us consider the action  \cite{Caprini:2014mja}~\footnote{We use the following notation:  $F_{\mu \nu} \equiv \partial_\mu A_\nu - \partial_\nu A_\mu$ is the field strength of $A_\mu$, while  $\tilde{F}^{\mu \nu} = \frac{ \eta^{\mu \nu \alpha \beta}}{2 \sqrt{-g}} F_{\alpha \beta}$ is the dual tensor, with $\eta^{0123} = 1$. We also use $M_p = 1/\sqrt{8 \pi G}$, where $G$ is Newton's constant.
In the following, dots (primes) denote derivatives with respect to physical (conformal) time, while $H \equiv \frac{\dot{a}}{a}$, where $a$ is the scale factor. }
\begin{eqnarray}
S =  \int d^4 x \sqrt{-g} 
\left[ \frac{M_p^2}{2} R - \frac{1}{2}\partial_\mu \phi \partial^\mu \phi
- V(\phi) 
+ I^2(\phi) \left( - \frac{1}{4}  F^{\mu \nu} F_{\mu \nu} 
  + \frac{\gamma}{4} \tilde{F}^{\mu \nu} F_{\mu \nu}
\right) \right] \, ,\label{eq:action}
\end{eqnarray}
with the parameter $\gamma$ being a constant. Differently from ref.~\cite{Caprini:2014mja}, we identify $\phi$ with the inflaton field, and compute how the couplings in eq.~\eqref{eq:action} affect the primordial perturbations. 
 
For $\gamma \neq 0$, the coupling of the inflaton field to the vector explicitly breaks parity, as the product $F^2$ is a scalar, while $F {\tilde F}$ is a pseudo-scalar quantity. This coupling also affects the inflaton perturbations, since expanding $\phi = \phi_0(\tau) + \delta \phi(\tau, {\bf x})$, and  denoting  $I(\phi_0) \equiv I_0(\tau)$, one has 
\begin{eqnarray}
I(\phi) = I_0
+ \frac{I_0'}{\phi_0'} \delta \phi 
\equiv I_0(\tau) + \delta I(\tau, {\bf x})\, .
\end{eqnarray}
In principle, ${\rm O} (\delta \phi^2)$ terms in the expansion of $I (\phi)$ could be considered, and included in the computation of the bispectrum. As discussed in ref.~\cite{Bartolo:2012sd} (see also ref.~\cite{Funakoshi:2012ym} for an explicit check), these higher terms give a subdominant contribution, and we can therefore disregard them. 

We assume that the function $I$ evolves in time during inflation as 
\begin{equation}
I_0(\tau) \propto a^n ( \tau) \;, 
\label{eq:n-def}
\end{equation}
and it then sets to a constant after inflation (when $\phi$ sets to a minimum).~\footnote{As we mentioned, the required time dependence can be obtained by suitably relating $I$ with the inflaton potential. See  ref.~\cite{Martin:2007ue} for details.} Without loss of generality, we can take this value to be $1$. Strong coupling considerations put the   $n > 0$ regime into question  \cite{Demozzi:2009fu}.  The coupling of the vector to any field charged under the U(1) symmetry is  $\propto \frac{1}{I_0}$. For $n>0$, this implies a very large coupling during inflation. Even if no real quanta of the charged particles exist during inflation, loop of virtual charged particles are out of perturbative control, which puts in question  any perturbative result obtained from the model  \cite{Demozzi:2009fu}.


\subsection{Background evolution}

A non-vanishing vev of the vector field during inflation leads to anisotropic expansion. It is well known  (see for instance ref.~\cite{Bartolo:2012sd}) that, for $\gamma = 0$, consistency with the  CMB results requires that  the energy density of the vector field is much smaller than that of $\phi$. We show below that the limit on the vector field energy becomes even stronger for $\gamma \neq 0$. Therefore, we can neglect the departure of the background geometry from the FLRW metric, and we use $ds^2 = a^2 (-d \tau^2 + dx^2)$ in our computations \cite{Bartolo:2012sd}. As in the standard case,  inflation is supported by the inflaton potential,  and the standard slow-roll  condition applies $\epsilon \equiv \frac{M_p^2}{2} \, \left( \frac{1}{V} \, \frac{d V}{d \phi} \right)^2 \simeq \frac{1}{2} \left( \frac{\dot{\phi}}{H M_p} \right)^2  \ll 1$, with negligible corrections. At zeroth order in slow roll, the dependence of the scale factor on conformal time is  $a \simeq -(H\tau)^{-1}$. 

We do not need to identify the vector field with the standard model photon. We nonetheless use the ``electromagnetic'' convention with
\begin{eqnarray}
{\bf E} = - \frac{I_0(\tau)}{a^2} {\bf A}' ~, \ \ 
{\bf B} = \frac{I_0(\tau)}{a^2} \nabla \times {\bf A} ~,
\end{eqnarray}
where the Coulomb gauge $A_0 = {\bf \nabla} \cdot {\bf A} = 0$ has been assumed. In this notation, the energy density in the gauge field acquires the familiar form $\rho_{\rm A} = \frac{{\bf E}^2+{\bf B}^2}{2}$. For convenience, we introduce the canonical field ${\bf V} \equiv I_0(\tau) {\bf A}$, and we expand it in a background value plus fluctuations as  ${\bf V} = {\bf V}^{(0)}(\tau) + \delta {\bf V}(\tau, {\bf x})$ (consistently with  spatial homogeneity, we have imposed that the background value depends on time only; we note that with this choice the term proportional to $\gamma$ in eq.~\eqref{eq:action} does not contribute to the background evolution). The gauge field vev then satisfies 
\begin{eqnarray}
\left({\bf V}^{(0)}\right)'' - \frac{I_0''}{I_0} {\bf V}^{(0)} = 0 ~,
\end{eqnarray}
leading to vanishing magnetic component, ${\bf B}^{(0)} = 0$, and to an electric component  ${\bf E}^{(0)}$ that depends on the parameter $n$.  In particular, for $n = -2$ (or $\frac{I_0''}{I_0} = \frac{2}{\tau^2}$), a time-independent vev arises as
\begin{eqnarray}
\label{eq:const_vev}
{\bf E}^{(0)}(\tau) = {\bf E}^{\rm vev}\, .
\end{eqnarray}
This is the case that we study in this work.

\subsection{Gauge field fluctuations}

The fluctuations of the gauge field have two helicity states ($\lambda = \pm 1$):~\footnote{The polarization vector $\epsilon_i^{(\lambda)}({\bf k})$ satisfies 
$\hat{k}_i \epsilon_i^{(\lambda)}(\hat{\bf k}) = 0$, 
$\eta^{0ijk} \hat{k}_i \epsilon_j^{(\lambda)}(\hat{\bf k}) = - \lambda i \epsilon_k^{(\lambda)}(\hat{\bf k})$, 
$\epsilon^{(\lambda) *}_i (\hat{\bf k}) = \epsilon^{(- \lambda)}_i (\hat{\bf k}) = \epsilon^{(\lambda)}_i (-\hat{\bf k})$ 
and $\epsilon^{(\lambda)}_i (\hat{\bf k}) \epsilon^{(\lambda')}_i (\hat{\bf k}) = \delta_{\lambda, -\lambda'}$. 
}
\begin{eqnarray}
 \delta V_i(\tau, {\bf x}) 
= \int \frac{d^3{\bf k}}{(2\pi)^{3/2}}
\sum_{\lambda = \pm 1} 
\delta {\hat V}_{\bf k}^{(\lambda)}(\tau)
\epsilon_i^{(\lambda)}({\bf k})
 {\rm e}^{i {\bf k} \cdot {\bf x}} ~,  
\end{eqnarray}
where the quantum field $\delta {\hat V}$ is decomposed as 
\begin{eqnarray}
\delta \hat{V}_{\bf k}^{(\lambda)}(\tau) 
= a_{\lambda}({\bf k}) \delta V_\lambda (\tau, k) 
+ a_{\lambda}^\dagger(- {\bf k}) \delta V_\lambda^* (\tau, k)  ~, 
\end{eqnarray}
in terms of creation and annihilation operators that obey the algebra  $[a_{\lambda}({\bf k}), a_{\lambda'}^\dagger({\bf k'})] = \delta_{\lambda \lambda'} \delta^{(3)}({\bf k} - {\bf k'})$. The quadratic action gives the evolution equation for the mode functions: 
\begin{eqnarray}
\delta V_\lambda''  
+ \left( k^2  + 2 \lambda k \gamma \frac{I_0'}{I_0}  - \frac{I_0''}{I_0} \right)
\delta V_\lambda 
= 0 ~.
\end{eqnarray}
The parity violating term results in a contribution that differs in sign for the two helicities.  Following ref.~\cite{Caprini:2014mja}, we define the  coupling strength parameter
\begin{eqnarray}
\xi \equiv -n \gamma \,, 
\end{eqnarray}
and for definiteness we assume $\xi > 0$. As we are interested in $n=-2$, this means that the parameter $\gamma $ is positive. In the opposite case, one simply needs to interchange $\delta V_+ \leftrightarrow \delta V_- $ in the results below.  Given that $I_0 \propto a^n$, one finds, at zeroth order in slow roll, 
\begin{eqnarray}
\delta V_\lambda''  
+ \left( k^2  + 2 \lambda k \frac{\xi}{\tau} - \frac{n \left( n + 1 \right)}{\tau^2} \right) \, 
\delta V_\lambda 
= 0 ~.
\label{eq:eom-dV}
\end{eqnarray}

Ref.~\cite{Caprini:2014mja} provided the solution to this equation with the standard adiabatic vacuum initial condition for arbitrary $n$. They obtained a particularly simple expression for  $\xi \gg 1$, and in the long wavelength regime. For $n=-2$, this expression reads 
\begin{eqnarray}
\delta V_+(\tau, k) \simeq - \frac{{\rm e}^{\pi \xi}}{\xi^{3/2}} \frac{\tau^{-1}}{2 \sqrt{\pi} k^{3/2}}  \;\;\;,\;\;\; 
\vert k \tau \vert \ll \frac{1}{\xi} \ll 1 \;, 
\label{eq:dV-p}
\end{eqnarray}
while the negative helicity mode $ \delta V_-(\tau, k) $ is produced on a much smaller and negligible amount ~\cite{Caprini:2014mja}. Using electromagnetic convention, we can write the following power spectra in the long wavelength regime 
\begin{eqnarray}
\Braket{\delta X_i(\tau, {\bf k}) \delta Y_j(\tau', {\bf k'})} 
\approx \delta X_+(\tau, k) \delta Y_+(\tau', k)
 \epsilon_i^{(+)}(\hat{\bf k}) \epsilon_j^{(+)}(\hat{\bf k}') 
\delta^{(3)}({\bf k} + {\bf k'})~, \label{eq:corr_E}
\end{eqnarray} 
where $X, Y = E, B$ and 
\begin{eqnarray}
\delta E_+(\tau, k) = 
- \frac{{\rm e}^{\pi  \xi}}{\xi^{3/2}}
\frac{3 H^2}{2\sqrt{\pi} k^{3/2}} 
= \frac{3}{k\tau} \delta B_+(\tau, k) ~.
\label{eq:super-dEdB}
\end{eqnarray}

This expression indicates that, on superhorizon scales ($-k\tau \ll 1$), the magnetic correlators are much smaller than  the electric ones. For this reason, we may ignore the magnetic-mode contributions in the primordial curvature and tensor correlations computed in section~\ref{sec:correlator}.

\subsection{Curvature and tensor perturbations}

To compute metric perturbations, we decompose them into scalar and tensor perturbations as standard \cite{Mukhanov:1990me}, and we work in the spatially flat gauge for the scalar sector, $\delta g_{ij}^{\rm scalar} = 0$. 
In this gauge, the spatial part of the metric fluctuation is given by the gravitational wave alone as $\delta g_{ij} = a^2 h_{ij}$. Moreover, since the energy density in the vector field is much smaller than that in the inflaton, we can identify the scalar curvature perturbation $\zeta \equiv - \frac{H \, \delta \rho}{\dot{\rho}} \simeq  - \frac{H}{\dot{\phi}} \delta\phi$, while the gauge field perturbations $\delta A_i$ are isocurvature modes that are produced as discussed in the previous subsection, and will affect $\zeta$ and $h_{ij}$ through their couplings to them.
\footnote{In the spatially flat gauge, there are also metric perturbations $\delta g_{00}$ and  $\delta g_{0i}$. These are non-dynamical modes that are integrated out, and induce gravitational couplings between the gauge quanta and $\zeta$. As the term proportional to $\gamma$ is a topological term, no metric perturbations enter there, so these couplings originate from the $I^2 F^2$ term, and are given in ref.~\cite{Barnaby:2012tk}. Such couplings are suppressed (technically, by an $\epsilon$ factor \cite{Barnaby:2012tk}) with respect to the direct $\delta \phi \delta A^2$ couplings present in eq.~\eqref{eq:action}, and so we can simply disregard them, and set $\delta g_{00} = \delta g_{0i} =0$ \cite{Barnaby:2012tk, Bartolo:2012sd,Bartolo:2014hwa}.}

We have already discussed the quantization of the $\delta A_i$ modes. The curvature and tensor perturbations are decomposed, respectively, as
\begin{eqnarray}
\zeta(\tau, {\bf x}) &=& \int \frac{d^3{\bf k}}{(2\pi)^{3/2}}
\zeta_{\bf k}(\tau) 
 {\rm e}^{i {\bf k} \cdot {\bf x}} ~, \\
h_{ij}(\tau, {\bf x}) &=& 
\int \frac{d^3{\bf k}}{(2\pi)^{3/2}}
\sum_{\lambda = \pm 2} 
h_{\bf k}^{(\lambda)}(\tau) 
e_{ij}^{(\lambda)}({\bf k})
 {\rm e}^{i {\bf k} \cdot {\bf x}} ~, 
\end{eqnarray}
where we have used the polarization tensor given by a product of the polarization vector as $e_{ij}^{(\lambda)}({\bf k}) \equiv \sqrt{2} \epsilon_i^{(\frac{\lambda}{2})}({\bf k}) \epsilon_j^{(\frac{\lambda}{2})}({\bf k})$. The quantized fields are expressed with the operators satisfying $[a_{\lambda}({\bf k}), a_{\lambda'}^\dagger({\bf k'})] = \delta_{\lambda \lambda'} \delta^{(3)}\delta({\bf k} - {\bf k'})$, respectively, as
\begin{eqnarray}
\hat{\zeta}_{\bf k}(\tau) 
&=& a_{0}({\bf k}) \zeta(\tau, k) 
+ a_{0}^\dagger(- {\bf k}) \zeta^* (\tau, k) ~, \label{eq:zeta_mode} \\
\hat{h}_{\bf k}^{(\lambda)}(\tau) 
&=& a_{\lambda}({\bf k}) h (\tau, k) 
+ a_{\lambda}^\dagger(- {\bf k}) h^* (\tau, k) ~. \label{eq:h_mode}
\end{eqnarray} 

We treat the effect of the gauge perturbations perturbatively, and indicate the scalar and tensor correlators as 
\begin{eqnarray} 
\begin{split}
  \Braket{{\hat \zeta}_{{\bf k}_1} {\hat \zeta}_{{\bf k}_2} } 
&= \Braket{{\hat \zeta}_{{\bf k}_1} {\hat \zeta}_{{\bf k}_2}}_0 + \Braket{{\hat \zeta}_{{\bf k}_1} {\hat \zeta}_{{\bf k}_2}}_1 ~, \\
  \Braket{{\hat \zeta}_{{\bf k}_1} {\hat \zeta}_{{\bf k}_2}  {\hat \zeta}_{{\bf k}_3} } &= \Braket{{\hat \zeta}_{{\bf k}_1} {\hat \zeta}_{{\bf k}_2}  {\hat \zeta}_{{\bf k}_3} }_0 + \Braket{{\hat \zeta}_{{\bf k}_1} {\hat \zeta}_{{\bf k}_2}  {\hat \zeta}_{{\bf k}_3} }_1 ~, \\
  \Braket{{\hat h}_{{\bf k}_1}^{(\lambda_1)} {\hat h}_{{\bf k}_2}^{(\lambda_2)}} 
&= \Braket{{\hat h}_{{\bf k}_1}^{(\lambda_1)} {\hat h}_{{\bf k}_2}^{(\lambda_2)}}_0 + \Braket{{\hat h}_{{\bf k}_1}^{(\lambda_1)} {\hat h}_{{\bf k}_2}^{(\lambda_2)}}_1  ~, \\ 
  \Braket{{\hat \zeta}_{{\bf k}_1} {\hat h}_{{\bf k}_2}^{(\lambda_2)}} 
&=  \Braket{{\hat \zeta}_{{\bf k}_1}  {\hat h}_{{\bf k}_2}^{(\lambda_2)}}_1 ~.
\end{split}
\label{eq:correlators}
\end{eqnarray}
The suffix zero refers to the correlators at zeroth-order in the gauge fields, namely to the standard inflationary vacuum correlators. We disregard the zeroth-order scalar three point function, as it would correspond to the non-Gaussianity from standard single-field slow-roll inflation, which is unobservable~\cite{Salopek:1990jq, Gangui:1993tt, Acquaviva:2002ud, Maldacena:2002vr}.

In absence of gauge fields, we have the standard mode function solutions 
\begin{eqnarray}
\zeta(\tau, k) = \frac{h(\tau, k)}{2 \sqrt{\epsilon}} 
= \frac{i H (1 + ik\tau)}{2\sqrt{\epsilon} M_p k^{3/2}} {\rm e}^{-ik\tau} \;\;,\;\; {\rm if } \; \delta A_i = 0 \;, 
\label{eq:0-mode}
\end{eqnarray}
leading to the standard power spectra 
\begin{eqnarray}
  \Braket{{\hat \zeta}_{{\bf k}_1} {\hat \zeta}_{{\bf k}_2} }_0 &=& 
\frac{2\pi^2}{k_1^3} \: {\cal P} \: \delta^{(3)}({\bf k}_1 + {\bf k}_2) \;\;\;,\;\;\; {\cal P} = \frac{H^2}{8\pi^2 \epsilon M_p^2} \;, 
\label{eq:zeta_zeta_0} \\
  \Braket{{\hat h}_{{\bf k}_1}^{(\lambda_1)} {\hat h}_{{\bf k}_2}^{(\lambda_2)}}_0
&=& \frac{8\pi^2}{k_1^3} \: \epsilon \: {\cal P} \: 
\delta^{(3)}({\bf k}_1 + {\bf k}_2) \: \delta_{\lambda_1, \lambda_2} 
 ~.  
\end{eqnarray}
 
The suffix $1$ in eq.~\eqref{eq:correlators} denotes the first non-vanishing correction from the vector fields. This requires two interaction terms, that we compute at tree level in the in-in formalism (see, e.g. ref.~\cite{Weinberg:2005vy}) in the next section.

\section{Correlators with broken parity and rotational symmetry}\label{sec:correlator}

In this section, we compute contributions from the gauge field to the correlators \eqref{eq:correlators}. As we mentioned, we treat the gauge field contributions perturbatively. The dominant coupling relevant for the scalar modes is obtained from the direct coupling $I^2 ( -\frac{1}{4}F^2 + \frac{\gamma}{4} F {\tilde F} )$. By expanding  $\delta (I^2(\phi)) \simeq - 2 \frac{I_0 I_0'}{aH} \zeta$, this gives rise to a term $\propto \zeta [ A^{(0)} + \delta A ]^2$ (as we discussed above, higher order expansions of $I^2$ and couplings that originate from $\delta g_{00}$ and $\delta g_{0i}$ can be neglected). From this expansion we have the two terms~\footnote{We note that the $A^{(0)\,2} \zeta$ tadpole is canceled once the exact equation of motion for $\zeta$ is taken into account, and the solution \eqref{eq:0-mode} is properly modified. This effect is completely negligible, due to the fact that $\rho_E^{\rm vev} \ll \rho_\phi$.} 
\begin{eqnarray}
S_{\zeta 1} 
&=& \int d\tau d^3 {\bf x}  \left(- \frac{2 a^3}{H} \right) \frac{I_0'}{I_0} \zeta 
\left[ 
\left(  E_i^{(0)} \delta E_i - B_i^{(0)} \delta B_i  \right) 
- \gamma \left(E_i^{(0)} \delta B_i + B_i^{(0)} \delta E_i \right)
\right]~, \label{eq:S_zetaAA} \\ 
S_{\zeta 2} 
&=&
  \int d \tau d^3 {\bf x} \left(- \frac{2 a^3}{H}\right) \frac{I_0'}{I_0} \zeta 
\left[
 \frac{1}{2}
\left( \delta E_i \delta E_i -  \delta B_i \delta B_i
 \right)
-  \gamma  \delta E_i \delta B_i
\right]~.
\end{eqnarray}

The $I^2 F^2$ term also leads to the dominant coupling between the gravity tensor mode and the gauge fields 
\begin{eqnarray}
S_{h 1} = 
 - \int d\tau d^3 {\bf x} a^4 h_{ij}
\left( E_i^{(0)} \delta E_j  + B_i^{(0)} \delta B_j \right) ~,
\end{eqnarray}
where we disregard the $h \, \delta A^2$ coupling as it does not contribute at tree level to two point correlators involving the graviton. 

In the constant vev case, i.e., $I_0 \propto a^{-2} = (-H \tau)^2$, these three terms give rise to the three interaction hamiltonians 
\begin{eqnarray}
H_{\zeta 1}(\tau) &=&  - \frac{4 \,E_i^{\rm vev}}{H^4 \, \tau^4}  
\int d^3 {\bf p}  \, \delta E_i(\tau, {\bf p}) \, \hat{\zeta}_{- \bf p}(\tau)  
~, \label{eq:H_zeta1} \\
H_{\zeta 2}(\tau) 
&=& 
 - \frac{2}{H^4 \, \tau^4} \int \frac{d^3 {\bf p} d^3 {\bf p}'}{(2\pi)^{3/2}} \, 
 \delta E_i(\tau, {\bf p}) \, 
\delta E_i(\tau, {\bf p}') \, 
\hat{\zeta}_{- {\bf p} - {\bf p}'}(\tau) ~, \label{eq:H_zeta2} \\ 
H_{h1}(\tau) 
&=& \frac{E_i^{\rm vev}}{H^4 \tau^4} 
\int d^3 {\bf p} \, \delta E_j(\tau, {\bf p}) 
\hat{h}_{ij, - {\bf p}}(\tau)  ~.
\end{eqnarray}

In $H_{\zeta 1}$ and $H_{\zeta 2}$, we have dropped the contribution of $\delta B_i$ with respect to $\delta E_i$, as the magnetic perturbation is much smaller than the electric one at super-horizon scales, see eq.~\eqref{eq:super-dEdB} (the super-horizon regime dominates the time integrals of the in-in computation  \cite{Bartolo:2012sd}, as we discuss below). 
Since $B^{(0)}_i =0$, and since $ F {\tilde F} \propto  E_i \, B_{i}$, this implies that the pseudo-scalar interaction  does not contribute to the leading order expression of  $H_{\zeta 1}$ and $H_{\zeta 2}$. For this reason, we recover the same dominant interaction hamiltonians as in ref.~\cite{Bartolo:2012sd}. Nonetheless, the $I^2 \left( \phi \right)  F {\tilde F}$ coupling strongly influences the phenomenological results, as it changes the gauge field mode functions, see eq.~\eqref{eq:super-dEdB}.

In the following, we perform the explicit computations of the correlators using the in-in formalism. A useful intermediate result is the commutator between the zeroth-order fields. Using the zeroth-order mode functions \eqref{eq:0-mode}, one finds 
\begin{eqnarray} 
\begin{split}
\Braket{ \left[ \hat{\zeta}_{{\bf k}_1}(\tau_1), \hat{\zeta}_{{\bf k}_2}(\tau_2) \right] }
 & \approx  
- \frac{i H^2}{6\epsilon M_p^2} 
\left[ \tau_1^3 - \tau_2^3 \right] \delta^{(3)}({\bf k}_1 + {\bf k}_2) 
~, \\
\Braket{ \left[ \hat{h}^{(\lambda_1)}_{{\bf k}_1}(\tau_1), \hat{h}^{(\lambda_2)}_{{\bf k}_2}(\tau_2) \right] }
& \approx  
- \frac{2 i H^2}{3 M_p^2} 
\left[ \tau_1^3 - \tau_2^3 \right] \delta^{(3)}({\bf k}_1 + {\bf k}_2) \delta_{\lambda_1, \lambda_2} 
 ~.
\end{split} \label{eq:corr_zeta_h}
\end{eqnarray}

\subsection{Power spectra}

By means of the in-in formalism, the leading order correction due to the gauge field to the two point scalar correlation function is given by 
\begin{eqnarray}
 \Braket{\prod_{n=1}^2 {\hat \zeta}_{{\bf k}_n}(\tau)}_{1} 
&=& - \int^{\tau} d\tau_1
\int^{\tau_1} d\tau_2
\Braket{ \left[\left[ \prod_{n=1}^2 \hat{\zeta}_{{\bf k}_n}(\tau), H_{\zeta 1}(\tau_1) \right], H_{\zeta 1}(\tau_2) \right]
}~. 
\label{eq:inin-zeta2}
\end{eqnarray}
We are interested in this result at super horizon scales,  $- k_n \tau \ll 1$. Modes of $\zeta$ and of $\delta E$ appearing in this expression are the zeroth-order solutions, and so they are uncorrelated with each other. So, the expectation value in eq.~\eqref{eq:inin-zeta2} splits into two separate expectation values. The $d \tau_i$ time integrals are dominated by modes in the long-wavelength regimes, since the small-wavelength modes are highly oscillatory, giving a highly suppressed contribution to the time integral \cite{Bartolo:2012sd}. On superhorizon scales, the electric field (contained in $H_{\zeta 1}(\tau_n)$) becomes a classical (commuting) field  \cite{Bartolo:2012sd}. Therefore, the expression \eqref{eq:inin-zeta2} is evaluated to 

\begin{eqnarray}
 \Braket{\prod_{n=1}^2 {\hat \zeta}_{{\bf k}_n}}_{1} 
&=& - \left( \frac{4 E_{\rm vev}}{H^4} \right)^2 
\int^{\tau} \frac{d\tau_1}{\tau_1^4} 
\int^{\tau_1} \frac{d\tau_2}{\tau_2^4} 
\int d^3 {\bf p}_1 \int d^3 {\bf p}_2  \nonumber \\ 
&& \hat{E}_i^{\rm vev} \hat{E}_j^{\rm vev} 
\Braket{ \delta E_i(\tau_1, {\bf p}_1) \delta E_j(\tau_2, {\bf p}_2) }
\nonumber \\
&&
\left( 
\Braket{ \left[ \hat{\zeta}_{{\bf k}_1}(\tau) , \hat{\zeta}_{- {\bf p}_1}(\tau_1)  \right] } 
\Braket{ \left[ \hat{\zeta}_{{\bf k}_2}(\tau) , \hat{\zeta}_{- {\bf p}_2}(\tau_2) \right] } 
+ 
({\bf k}_1 \leftrightarrow {\bf k}_2)
\right)
~.
\end{eqnarray}
After computing this with eqs.~\eqref{eq:corr_E} and \eqref{eq:corr_zeta_h}, the time integrals are reduced to $- \int_{-k_1^{-1}}^{\tau_{\rm e}} \frac{d\tau_i}{\tau_i}$, which is equivalent to the number of e-folds before the end of inflation at which the modes with $k_1$ leave the horizon, given by $N_{k_1}$. With an identity: 
\begin{eqnarray}
&& \left[ \epsilon_i^{(+)} ( \hat{\bf k}_1 )  \epsilon_j^{(+)} ( - \hat{\bf k}_1 ) + 
 \epsilon_i^{(+)} ( \hat{\bf k}_2 )  \epsilon_j^{(+)} ( - \hat{\bf k}_2 ) \right]  \hat{E}_i^{\rm vev} \hat{E}_j^{\rm vev}  \delta^{(3)} \left( {\bf k}_1 + {\bf k}_2 \right) \nonumber\\ 
&& = 
 \sum_{s= \pm 1} 
\epsilon_i^{(s)}(\hat{\bf k}_1) \epsilon_j^{(-s)}(\hat{\bf k}_1)
 \hat{E}_i^{\rm vev} \hat{E}_j^{\rm vev}  \delta^{(3)} \left( {\bf k}_1 + {\bf k}_2 \right) \nonumber\\ 
&& = \left[ 1 - \left(\hat{\bf k}_1 \cdot \hat{\bf E}^{\rm vev}\right)^2  \right]  \delta^{(3)} \left( {\bf k}_1 + {\bf k}_2 \right) 
~, 
\label{eq:polarization-id-gs}
\end{eqnarray}
we finally obtain 
\begin{eqnarray}
\Braket{\prod_{n=1}^2 {\hat \zeta}_{{\bf k}_n}}_{1} 
=
\frac{E_{\rm vev}^2}{2 \pi \epsilon^2 M_p^4} 
\frac{{\rm e}^{2 \pi \xi}}{\xi^{3}} \frac{N_{k_1}^2}{k_1^{3} }
\left[ 1 - \left(\hat{\bf k}_1 \cdot \hat{\bf E}^{\rm vev}\right)^2  \right]
\delta^{(3)}\left( {\bf k}_1 + {\bf k}_2 \right) \;\;\;,\;\;\; \xi \gg 1 \;,
 \label{eq:zeta_zeta}
\end{eqnarray}
which indeed shows a non-vanishing quadrupolar asymmetry, and which is scale invariant up to the logarithmic dependence on $k$ of $N_k$. 

It is instructive to compare this result to that obtained for $\xi = 2 \gamma = 0$  \cite{Bartolo:2012sd}. We find 
\begin{eqnarray}
\frac{  \Braket{ {\hat \zeta}_{{\bf k}_1} {\hat \zeta}_{{\bf k}_2}}_{1} \Big\vert_{\xi \gg 1}
}{   \Braket{{\hat \zeta}_{{\bf k}_1} {\hat \zeta}_{{\bf k}_2}}_{1}  \Big\vert_{\xi = 0}} = \frac{1}{4 \pi} \, \frac{{\rm e}^{2 \pi \xi}}{\xi^3} 
= \frac{\left( \delta E_+ \Big\vert_{\xi \gg 1} \right)^2}{\left( \delta E_+ \Big\vert_{\xi = 0 } \right)^2 + \left( \delta E_- \Big\vert_{\xi = 0 } \right)^2} \;. 
\end{eqnarray}
In short the difference between our result \eqref{eq:zeta_zeta} and the result obtained for $\gamma = \xi = 0$ is simply due to the difference between the wave functions of the sourcing gauge fields (the mode function in the present case is given in eq.~\eqref{eq:super-dEdB}; for $\xi =0$ one has instead $ \delta E_+  =  \delta E_-  = \frac{3 H^2}{\sqrt{2} k^{3/2}}$  \cite{Bartolo:2012sd}). 

By an analogous computation, the leading gauge field contributions to the scalar-tensor and to the tensor-tensor correlators are found to be 
\begin{eqnarray}
\Braket{{\hat \zeta}_{{\bf k}_1} {\hat h}_{{\bf k}_2}^{(\lambda_2)}}_{1} 
&=& 
- \frac{E_{\rm vev}^2}{\pi \epsilon M_p^4} 
\frac{{\rm e}^{2 \pi \xi}}{\xi^3} k_1^{-3} 
N_{k_1}^2
\hat{E}_i^{\rm vev} \hat{E}_j^{\rm vev} 
e_{ij}^{(\lambda_2)}(\hat{\bf k}_1)
\delta^{(3)}({\bf k}_1 + {\bf k}_2) \delta_{\lambda_2, 2} 
~, \label{eq:zeta_h} \\
\Braket{\prod_{n=1}^2 {\hat h}_{{\bf k}_n}^{(\lambda_n)}}_{1} 
&=& \frac{E_{\rm vev}^2}{\pi M_p^4} 
\frac{{\rm e}^{2 \pi  \xi}}{\xi^3} k_1^{-3}  N_{k_1}^2 
\left[ 1 - \left(\hat{\bf k}_1 \cdot \hat{\bf E}^{\rm vev}\right)^2  \right]
 \delta^{(3)}({\bf k}_1 + {\bf k}_2) \delta_{\lambda_1, 2} \delta_{\lambda_2, 2} ~. \label{eq:h_h}
\end{eqnarray}
The absence of the $\lambda = - 2$ mode in these expressions indicates parity-violating correlators. Moreover, these correlators also break isotropy due to non-vanishing quadrupolar term. The gauge field contributions to the scalar-tensor and tensor-tensor correlator are suppressed, respectively, by a factor  $\epsilon$ and by a factor $\epsilon^2$ with respect to the contribution to the scalar-scalar correlator. An analogous behavior was also found in the $F^2 + \phi F \tilde{F}$ model \cite{Bartolo:2014hwa}.

\subsection{Scalar bispectrum}

Again using the in-in formalism, the leading order contribution of the gauge field to the scalar bispectrum is given by :
\begin{eqnarray}
\Braket{\prod_{n=1}^3 {\hat \zeta}_{{\bf k}_n}(\tau)}_1
&=& \Braket{\prod_{n=1}^3 {\hat \zeta}_{{\bf k}_n}(\tau)}^{(211)} + 
\Braket{\prod_{n=1}^3 {\hat \zeta}_{{\bf k}_n}(\tau)}^{(121)} + 
\Braket{\prod_{n=1}^3 {\hat \zeta}_{{\bf k}_n}(\tau)}^{(112)}
~, 
\label{eq:inin-zeta3}
\end{eqnarray}
where we have defined 
\begin{eqnarray}
\Braket{\prod_{n=1}^3 {\hat \zeta}_{{\bf k}_n}(\tau)}^{(abc)} &\equiv& 
i \int^\tau d\tau_1  \int^{\tau_1} d\tau_2 \int^{\tau_2} d\tau_3  
\nonumber \\ 
&& 
\Braket{ \left[\left[\left[ \prod_{n=1}^3 \hat{\zeta}_{{\bf k}_n}(\tau), H_{\zeta a}(\tau_1) \right], H_{\zeta b}(\tau_2) \right], H_{\zeta c}(\tau_3) \right] 
} ~.
\end{eqnarray}

As for the power spectrum, the time integrals are dominated by modes in the super-horizon regime. Proceeding as in that computation, the first term in eq.~\eqref{eq:inin-zeta3} evaluates to 
\begin{eqnarray}
\Braket{\prod_{n=1}^3 {\hat \zeta}_{{\bf k}_n}}^{(211)}
&=& - i \frac{32 E_{\rm vev}^2}{H^{12}} \int^{\tau} \frac{d\tau_1}{\tau_1^4}  
\int^{\tau_1} \frac{d\tau_2}{\tau_2^4} \int^{\tau_2} \frac{d\tau_3}{\tau_3^4}  
 \int \frac{d^3 {\bf p}_1 d^3 {\bf p}_1'}{(2\pi)^{3/2}} 
\int d^3 {\bf p}_2  
\int d^3 {\bf p}_3 
 \nonumber \\ 
&& 
\hat{E}_j^{\rm vev} \hat{E}_k^{\rm vev} 
\Braket{ \delta E_i(\tau_1, {\bf p}_1) 
\delta E_i(\tau_1, {\bf p}_1') 
\delta E_j(\tau_2, {\bf p}_2)
 \delta E_k(\tau_3, {\bf p}_3) } \nonumber \\ 
&& 
\Braket{\left[\hat{\zeta}_{{\bf k}_1}(\tau), \hat{\zeta}_{- {\bf p}_1 - {\bf p}_1'}(\tau_1)\right]}
\Braket{\left[\hat{\zeta}_{{\bf k}_2}(\tau), \hat{\zeta}_{-{\bf p}_2}(\tau_2)\right]}
\Braket{\left[\hat{\zeta}_{{\bf k}_3}(\tau), \hat{\zeta}_{-{\bf p}_3}(\tau_3)\right]} \nonumber \\ 
&&+ 5~{\rm perms ~ in} ~ {\bf k}_n 
\end{eqnarray}
This expression evaluates to 
\begin{eqnarray}
\Braket{\prod_{n=1}^3 {\hat \zeta}_{{\bf k}_n}}^{(211)}
&\simeq& \frac{3 E_{\rm vev}^2 \, H^2}{2 \pi^2 \left( 2 \pi \right)^{3/2} \epsilon^3 M_p^6} \, 
\frac{\delta^{(3)} \left( {\bf k}_1 + {\bf k}_2 + {\bf k}_3 \right)}{k_2^3 \, k_3^3} \, \frac{{\rm e}^{4 \pi \xi}}{\xi^6} \nonumber\\ 
&& \epsilon_i^{(+)*}(\hat{\bf k}_2) \epsilon_j^{(+)}(\hat{\bf k}_2) \epsilon_i^{(+)*}(\hat{\bf k}_3) \epsilon_k^{(+)}(\hat{\bf k}_3) \hat{E}_j^{\rm vev} \hat{E}_k^{\rm vev} \nonumber\\ 
&& 
\int_{{\rm Max} \left[ - \frac{1}{k_1} ,\,  - \frac{1}{k_2} ,\,  - \frac{1}{k_3} \right]}^{\tau} d \tau_1 \frac{\tau^3 - \tau_1^3}{\tau_1^4} \, 
\int_{ - \frac{1}{k_2} }^{\tau_1} d \tau_2 \frac{\tau^3 - \tau_2^3}{\tau_2^4} \, 
\int_{ - \frac{1}{k_3} }^{\tau_1} d \tau_3 \frac{\tau^3 - \tau_3^3}{\tau_3^4} \nonumber\\ 
&&+ 2~{\rm perms ~ in} ~ {\bf k}_n \;\;\;,\;\;\; \xi \gg 1 \;. 
\label{eq:z3-par}
\end{eqnarray}

Namely, the internal time variables in the integrals cover the region $\tau \geq \tau_1 \geq \tau_2 ,\, \tau_3$, plus lower bounds dictated by the requirement that all the modes in the interaction are in the super-horizon regime. Starting from the other two terms in eq.~\eqref{eq:inin-zeta3}, and relabeling the internal times, we obtain an expression where the integrand is identical to the integrand of eq.~\eqref{eq:z3-par}, but the internal times cover the two complementary regions  $\tau \geq \tau_2 \geq \tau_3 ,\, \tau_1$, and  $\tau \geq \tau_3 \geq \tau_1 ,\, \tau_2$. This leads to  
\begin{eqnarray}
\Braket{\prod_{n=1}^3 {\hat \zeta}_{{\bf k}_n}}_1
&\simeq& \frac{3 E_{\rm vev}^2 \, H^2}{2 \pi^2 \left( 2 \pi \right)^{3/2} \epsilon^3 M_p^6} \, 
\frac{\delta^{(3)} \left( {\bf k}_1 + {\bf k}_2 + {\bf k}_3 \right)}{k_2^3 \, k_3^3} \, \frac{{\rm e}^{4 \pi \xi}}{\xi^6} \nonumber\\ 
&& \epsilon_i^{(+)*}(\hat{\bf k}_2) \epsilon_j^{(+)}(\hat{\bf k}_2) \epsilon_i^{(+)*}(\hat{\bf k}_3) \epsilon_k^{(+)}(\hat{\bf k}_3) \hat{E}_j^{\rm vev} \hat{E}_k^{\rm vev} \nonumber\\ 
&& 
\int_{{\rm Max} \left[ - \frac{1}{k_1} ,\,  - \frac{1}{k_2} ,\,  - \frac{1}{k_3} \right]}^\tau d \tau_1 \frac{\tau^3 - \tau_1^3}{\tau_1^4} \, 
\int_{ - \frac{1}{k_2} }^{\tau} d \tau_2 \frac{\tau^3 - \tau_2^3}{\tau_2^4} \, 
\int_{ - \frac{1}{k_3} }^{\tau} d \tau_3 \frac{\tau^3 - \tau_3^3}{\tau_3^4} \nonumber\\ 
&&+ 2~{\rm perms ~ in} ~ {\bf k}_n \;\;\;,\;\;\; \xi \gg 1 \;. 
\end{eqnarray}

The time integrations are dominated by the earlier possible times, giving 
\begin{eqnarray}
\Braket{\prod_{n=1}^3 {\hat \zeta}_{{\bf k}_n}}_1
&=& \frac{\delta^{(3)}({\bf k}_1 + {\bf k}_2 + {\bf k}_3)}{(2\pi)^{3/2}} 
{\cal C}_{\hat{\bf k}_1 ,\, \hat{\bf k}_2 ,\, \hat{\bf E}^{\rm vev}}^{\xi \gg 1} 
f(k_1, k_2, k_3 ;\, \xi) \nonumber\\ 
&& 
+ 2 ~{\rm perms ~ in} ~ {\bf k}_n \;\;\;,\;\;\; \xi \gg 1 \;.  
\label{eq:zeta_zeta_zeta}
\end{eqnarray}
We here have defined 
\begin{eqnarray}
{\cal C}_{\hat{\bf k}_1 ,\, \hat{\bf k}_2 ,\, \hat{\bf E}^{\rm vev}}^{\xi \gg 1} 
&\equiv& 
\epsilon_i^{(+)*}(\hat{\bf k}_1) \epsilon_j^{(+)}(\hat{\bf k}_1) 
\epsilon_i^{(+)*}(\hat{\bf k}_2) \epsilon_k^{(+)}(\hat{\bf k}_2)
\hat{E}_j^{\rm vev} \hat{E}_k^{\rm vev} ~, \\
f \left(k_1, k_2 , k_3 ;\, \xi \right)
&\equiv& \frac{3 E_{\rm vev}^2 H^2}{2 \pi^2 \epsilon^3 M_p^6} 
\frac{{\rm e}^{4 \pi  \xi}}{\xi^{6}}
 \frac{ {\rm Min}[N_{k_1}, N_{k_2} , N_{k_3} ] N_{k_1} N_{k_2} }{ k_1^3 \, k_2^3 }   \;\;\;,\;\;\; \xi \gg 1 \;, 
\end{eqnarray}
where we recall that $N_{k_i}$ is the number of e-fold before the end of inflation at which the mode with momentum $k_i$ left the horizon. 
Using the identity $\epsilon_i^{(+)*} ( \hat{\bf k} ) \epsilon_j^{(+)} 
(\hat{\bf k}) = \frac{1}{2} \left[ \delta_{ij} - \hat{k}_i \hat{k}_j + i \, \eta^{0ijk} \hat{k}_k \right]$, the angle dependence in ${\cal C}_{\hat{\bf k}_1,\,  \hat{\bf k}_2,\,  \hat{\bf E}^{\rm vev}}^{\xi \gg 1}$ can be simplified as
\begin{eqnarray} 
{\cal C}_{\hat{\bf k}_1 ,\, \hat{\bf k}_2 ,\, \hat{\bf E}^{\rm vev}}^{\xi \gg 1} 
&=& \frac{1}{4} 
\left\{ 1 - \left( \hat{\bf k}_1 \cdot \hat{\bf E}^{\rm vev} \right)^2  
- \left( \hat{\bf k}_2 \cdot \hat{\bf E}^{\rm vev} \right)^2 
+  \left( \hat{\bf k}_1 \cdot \hat{\bf E}^{\rm vev} \right) 
\left( \hat{\bf k}_2 \cdot \hat{\bf E}^{\rm vev} \right) 
\left( \hat{\bf k}_1 \cdot \hat{\bf k}_2 \right) \right. 
\nonumber\\ 
&&\left. \quad
- \, \hat{\bf k}_1 \cdot \hat{\bf k}_2 
+ \left( \hat{\bf k}_1 \cdot \hat{\bf E}^{\rm vev} \right) 
\left( \hat{\bf k}_2 \cdot \hat{\bf E}^{\rm vev} \right)  \right. 
\nonumber\\ 
&&\left. \quad
+ \, i \left[ \hat{\bf E}^{\rm vev} \cdot \left( \hat{\bf k}_1 - \hat{\bf k}_2 \right) \right] 
 \left[ \hat{\bf E}^{\rm vev} \cdot \left( \hat{\bf k}_1 \times \hat{\bf k}_2 \right) \right] \right\} ~.
\end{eqnarray}

In the $I^2(\phi) F^2$ model (the $\gamma = \xi = 0$ case), the corresponding contribution to the bispectrum is \cite{Bartolo:2012sd}
\begin{eqnarray}
\left. \Braket{\prod_{n=1}^3 {\hat \zeta}_{{\bf k}_n}}_1   \right\vert_{\xi = 0} 
&=&  
\frac{\delta^{(3)}({\bf k}_1 + {\bf k}_2 + {\bf k}_3)}{(2\pi)^{3/2}} 
{\cal C}_{\hat{\bf k}_1 ,\, \hat{\bf k}_2 ,\, \hat{\bf E}^{\rm vev}}^{\xi = 0} 
f  (k_1, k_2, k_3 ;\, \xi = 0) \nonumber \\ 
&&
+ 2~{\rm perm~in~} {\bf k}_n ~,
\end{eqnarray}
where
\begin{eqnarray}
{\cal C}_{\hat{\bf k}_1 ,\, \hat{\bf k}_2 ,\, \hat{\bf E}^{\rm vev}}^{\xi = 0} 
&=& \left[ \sum_{s_1 = \pm 1}  
\epsilon_i^{(s_1)}(\hat{\bf k}_1) \epsilon_j^{(s_1)*}(\hat{\bf k}_1) \right]
\left[ \sum_{s_2 = \pm 1}  
\epsilon_i^{(s_2)}(\hat{\bf k}_2) \epsilon_k^{(s_2)*}(\hat{\bf k}_2) \right] 
 \hat{E}_j^{\rm vev} 
 \hat{E}_k^{\rm vev} \nonumber \\ 
&=& 1 - \left( \hat{\bf k}_1 \cdot  \hat{\bf E}^{\rm vev}  \right)^2 
- \left( \hat{\bf k}_2 \cdot  \hat{\bf E}^{\rm vev}  \right)^2 \nonumber \\ 
&& + \left( \hat{\bf k}_1 \cdot \hat{\bf E}^{\rm vev} \right) 
\left( \hat{\bf k}_2 \cdot \hat{\bf E}^{\rm vev} \right) 
\left( \hat{\bf k}_1 \cdot \hat{\bf k}_2 \right)
~, \\ 
f \left(k_1, k_2,k_3 ;\, \xi =0 \right) &=&  f \left(k_1, k_2,k_3 ;\, \xi \gg 1 \right) \times \left( \frac{ \delta E \Big\vert_{\xi =0} }{  \delta E_+ \Big\vert_{\xi \gg 1} } \right)^4 \;,  
\end{eqnarray}
with $ \delta E \Big\vert_{\xi =0} = \frac{3 H^2}{\sqrt{2} k^{3/2}} $ being the mode function of either helicity mode in the $\xi=0$ case (where the two helicities are produced in equal amount). As for the power spectrum, the difference between the $\xi \gg 1$ and the $\xi =0$ result is simply due to the difference of the sourcing gauge fields. In the present case, one helicity dominates the final result, leading to violation of parity.

\section{CMB constraints}\label{sec:constraints}

In this section, we compare the primordial correlators induced by the action \eqref{eq:action} to the latest CMB data, and we obtain an upper bound on the energy density in the background gauge field, $\rho_E^{\rm vev}$. Finally, we compare this bound with the theoretically expected value for the vev.

\subsection{CMB observables and constraints} \label{subsec:observable}

The primordial correlators computed in section~\ref{sec:correlator} act as the initial conditions of the CMB correlators and can therefore be measured by CMB observations. Formally, the signatures in the CMB scalar-scalar, scalar-tensor and tensor-tensor power spectra are essentially identical to those obtained in the $F^2 + \phi F \tilde{F}$ model \cite{Bartolo:2014hwa}, since the primordial correlators have the identical angular dependence. The main difference between the signatures on that model and the signatures that we are computing here is that in the present case the gauge field vev is constant, so that these signatures are nearly scale invariant. On the contrary, in the  $F^2 + \phi F \tilde{F}$ case, the vector vev (that needs to be assumed as an ad hoc initial condition present when the CMB modes were generated) is rapidly decreasing, leading to observational effects only at the largest scales. 

As proven in ref.~\cite{Bartolo:2014hwa}, the scalar-scalar correlator \eqref{eq:zeta_zeta} can create TT, TE and EE in $|\ell_1 - \ell_2| = 0, 2$, while the scalar-tensor \eqref{eq:zeta_h} or tensor-tensor \eqref{eq:h_h} correlator can generate TT, TE, EE, BB, TB and EB in $|\ell_1 - \ell_2| = 0, 1, 2$. Specifically, TT, TE, EE and BB (TB and EB) in $|\ell_1 - \ell_2| = 1$ ($|\ell_1 - \ell_2| = 2$) are distinct signatures of the anisotropic pseudoscalar inflation, since these appear only in the case that parity and rotational symmetries are broken at the same time. However, these signal-to-noise ratios are smaller than the scalar-scalar ones because of the slow-roll suppression of the tensor mode. Likewise, the tensor mode also gives  special correlations in the bispectrum due to broken parity and rotational invariance; also such contributions are subdominant due to the smallness of the tensor mode. In this section, we therefore focus on the observables associated with the primordial scalar mode $\zeta$.

For this analysis, we use the conventional  $g_*$ parametrization for the power spectrum \cite{Ackerman:2007nb} and the $c_L$ parametrization for the bispectrum \cite{Shiraishi:2013vja}, namely, 
\begin{eqnarray}
\Braket{\prod_{n = 1}^2 \zeta_{{\bf k}_n}} 
&=& \delta^{(3)}\left( {\bf k}_1 + {\bf k}_2 \right) 
P(k_1)
\left[ 1 + g_* \left( \hat{\bf k}_1 \cdot \hat{\bf E}^{\rm vev} \right)^2  \right]
 ~, \label{eq:g_*_def} \\ 
\Braket{\prod_{n = 1}^3 \zeta_{{\bf k}_n}} 
&=& \frac{\delta^{(3)}\left( {\bf k}_1 + {\bf k}_2 +  {\bf k}_3 \right)}{(2\pi)^{3/2}}
\sum_{L} c_L P(k_1) P(k_2) P_L(\hat{\bf k}_1 \cdot \hat{\bf k}_2) +  
2~{\rm perms} ~, \label{eq:c_L_def}
\end{eqnarray}
where $P_L(x)$ is the Legendre polynomial. The latter bispectrum form is obtained after averaging the original anisotropic bispectrum \eqref{eq:zeta_zeta_zeta} over all directions of $\hat{\bf E}^{\rm vev}$, in the spirit of isotropic CMB measurements. This is the quantity that is immediately associated to the angle-averaged reduced bispectrum computed from the data: the reduced bispectrum $b \left( k_1 ,\, k_2 ,\, k_3 \right)$ is obtained by averaging the bispectrum over all possible orientation of triangles of sides of length $k_1 ,\, k_2 ,\, k_3$. The theoretical prediction for the reduced bispectrum associated to eq.~\eqref{eq:zeta_zeta_zeta} is therefore equivalent to the theoretical prediction associated to the average of eq.~\eqref{eq:zeta_zeta_zeta} over all possible direction of  $\hat{\bf E}^{\rm vev}$ \cite{Shiraishi:2013vja}. 
\footnote{
In the study of CMB anisotropic bispectra~\cite{Bartolo:2011ee,Shiraishi:2011ph} this would correspond to single-out a monopole term in a spherical harmonic expansion of the anisotropic bispectrum~\cite{Bartolo:2011ee} that does contribute to the isotropic (angle-averaged) bispectrum. 
}

Keeping into account that  the CMB  data force $\vert g_* \vert \ll 1$ \cite{Kim:2013gka, Ade:2015lrj},  the sum of eqs.~\eqref{eq:zeta_zeta_0} and \eqref{eq:zeta_zeta} yields $P(k) \simeq \frac{2\pi^2}{k^3} {\cal P} $ and 
\begin{eqnarray}
\xi \gg 1 \;:\;\;\;\; 
g_* \simeq - \frac{12 N_{\rm CMB}^2}{\pi \epsilon} 
\frac{{\rm e}^{2 \pi \xi}}{\xi^{3}} \frac{\rho_E^{\rm vev}}{\rho_\phi} \;, 
 \label{eq:g_*}
\end{eqnarray}
where $N_{\rm CMB}$ is the number of e-folds before the end of inflation at which the CMB modes leave the horizon,  $\rho_\phi \simeq V(\phi) \simeq 3 M_p^2 H^2$ is the energy density of the inflaton, and $\rho_E^{\rm vev} \equiv \frac{E_{\rm vev}^2}{2}$ is the energy density of  the gauge field. 

Let us now compute the average of the bispectrum \eqref{eq:zeta_zeta_zeta} over all directions of  $\hat{\bf E}^{\rm vev}$. Using~\footnote{This identity is easily derived using the spherical-harmonics representations of a unit vector and polarization vector, see \cite{Shiraishi:2010kd}.} 
\begin{eqnarray}
&& \int \frac{d^2 \hat{\bf E}^{\rm vev}}{4\pi} 
{\cal C}_{\hat{\bf k}_1 ,\, \hat{\bf k}_2 ,\, \hat{\bf E}^{\rm vev}}^{\xi \gg 1} 
 =  \frac{1}{9}   
P_{0}(\hat{\bf k}_1 \cdot \hat{\bf k}_2) 
- \frac{1}{6} 
P_{1}(\hat{\bf k}_1 \cdot \hat{\bf k}_2)
+ \frac{1}{18}  
P_{2}(\hat{\bf k}_1 \cdot \hat{\bf k}_2) ~, \label{eq:angle_average_pseudo}
\end{eqnarray}
and setting $N_{k_1} \simeq N_{k_2} \simeq N_{k_3} \simeq N_{\rm CMB}$, we obtain 
\begin{eqnarray}
\xi \gg 1 \;:\;\;\;\; 
 c_0 = - \frac{4N_{\rm CMB}}{3\pi} \frac{{\rm e}^{2 \pi \xi}}{\xi^{3}} g_* 
\;\;,\;\;  c_1 = - \frac{3 c_0}{2} \;\;,\;\; c_2 = \frac{c_0}{2} \;. 
\label{eq:c_L} 
\end{eqnarray}

We compare eqs.~\eqref{eq:g_*} and \eqref{eq:c_L} with the results obtained for the $\gamma = \xi = 0$ case  \cite{Shiraishi:2013vja} 
\begin{eqnarray}
\xi = 0 \;:\;\;\;\; 
 g_* \simeq - \frac{48 \, N_{\rm CMB}^2}{\epsilon} \, \frac{\rho_E^{\rm vev}}{\rho_\phi} 
\;\;,\;\; c_0 = - \frac{16}{3}  N_{\rm CMB} \, g_* \;\;,\;\; c_1 = 0 \;\;,\;\; c_2 = \frac{c_0}{2} \;. 
\label{eq:gs-c_L-g0} 
\end{eqnarray}

The {\it Planck} collaboration reported the $95\%$ CL limits \cite{Ade:2015lrj, Ade:2015ava}
\footnote{These $c_L$ bounds correspond to the temperature only limits \cite{Ade:2015ava}: $f_{\rm NL}^{\rm local} = 2.5 \pm 5.7$, $f_{\rm NL}^{L=1} = -49 \pm 43$ and $f_{\rm NL}^{L=2} = 0.5 \pm 2.7$ ($68\%$ CL), where $c_0 = (6/5) f_{\rm NL}^{\rm local}$, $c_1 = -(12/5) f_{\rm NL}^{L=1}$ and $c_2 = -(48/5) f_{\rm NL}^{L=2}$ hold.}
\begin{eqnarray}
- 0.0225 \leq g_* \leq 0.0363 \;, \;\;\; 
- 10.7 \leq c_0 \leq 16.7 \;, \;\;\; 
- 89 \leq c_1 \leq 324 \;, \;\;\; 
- 57 \leq c_2 \leq 47 \;.   
\label{eq:Planck-bounds}
\end{eqnarray}
From these constraints we get rough estimates of the upper bounds on $\rho_E^{\rm vev}/\rho_\phi$ using eqs.~\eqref{eq:g_*}, \eqref{eq:c_L} and \eqref{eq:gs-c_L-g0}:
\begin{eqnarray}
\gamma = 0 \;:\;\;\;\;  \frac{\rho_E^{\rm vev}}{\rho_\phi} \leq 
\begin{cases}
1.3 \times 10^{-9} \left( \frac{\epsilon}{0.01} \right) \,  \left( \frac{60}{N_{\rm CMB}} \right)^2  & ({\rm from} \ g_*) \;, \\
3.0 \times 10^{-9}  \,  \left( \frac{\epsilon}{0.01} \right)  \left( \frac{60}{N_{\rm CMB}} \right)^3  & ({\rm from} \ c_0) \;, \\
1.7 \times 10^{-8}   \,  \left( \frac{\epsilon}{0.01} \right)  \left( \frac{60}{N_{\rm CMB}} \right)^3  & ({\rm from} \ c_2) \;,
\end{cases}
\label{eq:bound1}
\end{eqnarray}
and
\begin{eqnarray}
\gamma \gg \frac{1}{2} \;:\;\;\;\; 
\frac{\rho_E^{\rm vev}}{\rho_\phi} &\leq& 
\begin{cases}
1.3 \times 10^{-7} \left(\frac{\gamma^3}{{\rm e}^{4 \pi \gamma}}\right) 
\,  \left( \frac{\epsilon}{0.01} \right) 
 \,  \left( \frac{60}{N_{\rm CMB}} \right)^2
& ({\rm from} \ g_*) \;, \\
3.1 \times 10^{-5} \left( \frac{\gamma^3}{{\rm e}^{4 \pi \gamma}} \right)^2 
\,  \left( \frac{\epsilon}{0.01} \right) 
 \,  \left( \frac{60}{N_{\rm CMB}} \right)^3
& ({\rm from} \ c_0) \;, \\
1.1 \times 10^{-4} \left( \frac{\gamma^3}{{\rm e}^{4 \pi \gamma}} \right)^2 
\,  \left( \frac{\epsilon}{0.01} \right) 
 \,  \left( \frac{60}{N_{\rm CMB}} \right)^3 
& ({\rm from} \ c_1) \;, \\
1.7 \times 10^{-4} \left( \frac{\gamma^3}{{\rm e}^{4 \pi \gamma}} \right)^2 
\,  \left( \frac{\epsilon}{0.01} \right) 
 \,  \left( \frac{60}{N_{\rm CMB}} \right)^3 
& ({\rm from} \ c_2) \;.
\end{cases}
\label{eq:bound2}
\end{eqnarray}

\begin{figure}[t]
  \begin{tabular}{c}
    \begin{minipage}{1.0\hsize}
  \begin{center}
    \includegraphics[width = 1.0\textwidth]{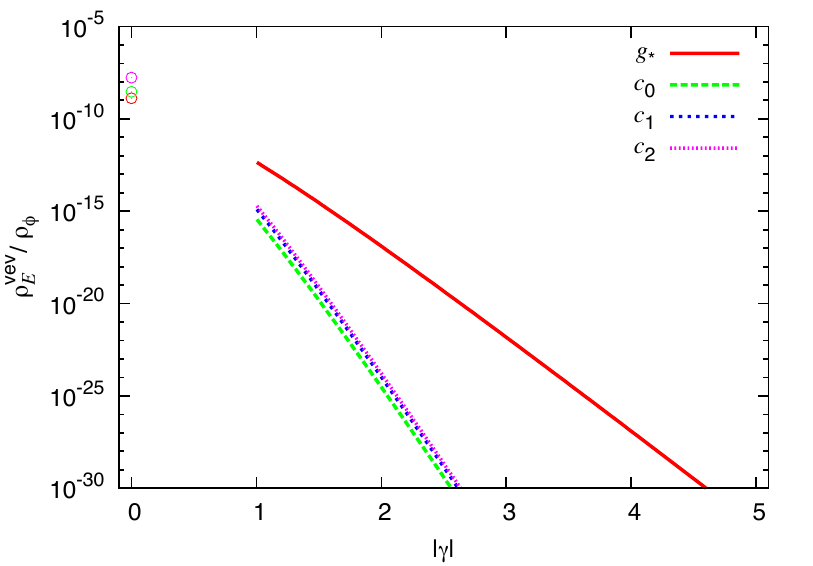}
  \end{center}
\end{minipage}
\end{tabular}
\caption{Upper bounds on $\rho_E^{\rm vev} / \rho_\phi$  reported in eqs.~\eqref{eq:bound1} and \eqref{eq:bound2},  with $N_{\rm CMB} = 60$ and $\epsilon = 0.01$. As all these bounds must be satisfied, the final bound on 
 $\rho_E^{\rm vev} / \rho_\phi$ is the most stringent of the bounds shown. Strictly speaking, the solid and dashed lines are based on an approximation that holds for $\vert \gamma \vert \gg \frac{1}{2}$ (that is, $\vert \xi \vert \gg 1$). We extended them up to $|\gamma| =1$, as a naive extrapolation of the lines shown in the figure is roughly consistent with the values for $\vert \gamma \vert =0$, that are exact.} 
\label{fig:95CL_gamma_Ephi}
\end{figure}

We show these bounds in figure \ref{fig:95CL_gamma_Ephi}, for the special choices of $N_{\rm CMB} = 60$ and $\epsilon = 0.01$. While at $\gamma=0$ the strongest bound on   $\rho_E^{\rm vev} / \rho_\phi$  is obtained from the power spectrum, at $\gamma \gg \frac{1}{2}$ the strongest bound is due to the bispectrum coefficient $c_0$. This is due the fact that the gauge production increases with increasing $\gamma$, and the bispectrum is affected more than the power spectrum by this growth (at the technical level, the source for the power spectrum is $\propto \delta E^2$, while the one for the bispectrum is  $\propto \delta E^4$. This generates the  ${\rm e}^{2 \pi \xi} / \xi^3$ factor in $c_0/g_*$ in eq.~\eqref{eq:c_L}). 

The result we have obtained assumed  $\xi = 2 \gamma > 0$, but it can be readily applied also to the case in which 
  $\xi = 2 \gamma < 0$. As clear from eq.~\eqref{eq:eom-dV}, changing sign of $\xi = 2 \gamma$ simply amounts in interchanging the role of the two helicities of the gauge field. In the computation of the inflation correlators, one simply needs to replace  $\xi \rightarrow \vert \xi \vert$ and interchange $\epsilon_i^{(+)} \, \leftrightarrow \epsilon_i^{(-)}$. One can then verify that the limits \eqref{eq:bound2} can be extended to the $\gamma \ll -\frac{1}{2}$ region by simply replacing $\gamma$ with $\vert \gamma \vert$.

  
\subsection{Comparison with the expected vector vev} \label{subsec:expected}  
 
In this subsection we compare the limits  on $\rho_E^{\rm vev} / \rho_\phi $ obtained in the previous subsection with what is naturally expected in the model \eqref{eq:action}.   

Beside producing scale invariant vector perturbations, the  $I \propto a^{-2}$ case also sustains a constant vector vev. This is not accidental. Due to scale invariance, super horizon modes have the same power at any given time $t$. The  modes also had the same power when they left the horizon.~\footnote{This is just a statement of time translation invariance during inflation; this invariance is broken by the slow roll parameters, as the inflaton typically speeds up during inflation. This leads to departure from scale invariance - typical in slow roll inflation - that we are disregarding in the present discussion.} Equal power at horizon crossing (that happens at different time for the different modes) and at the later common time $t$ is obtained because the modes have frozen amplitude while outside the horizon.   In the super-horizon regime a mode solves the same equation of motion as a constant vev. So, if the super-horizon modes are frozen outside the horizon, also a completely homogeneous vev is. This explains why the  $I \propto a^{-2}$ choice produces both scale invariant electric perturbations, and a constant electric vev. 

In presence of the vector vev, the interaction between the gauge and the inflaton perturbations leads to anisotropic correlators of the primordial perturbations. The fact that this anisotropy is not observed in the data results in a  limit on the vev of the vector field, and on the associated energy density $\rho_E^{\rm vev} \equiv \frac{{\bf E}_{\rm vev}^2}{2}$. The current limit is summarized in figure~\ref{fig:95CL_gamma_Ephi}. 

Naively, one may simply assume that the gauge field has no homogeneous vev, in which case the limit computed here does not apply. While this is a mathematical possibility, such an assumption is very unlikely in a model that is constructed to give 
scale invariant gauge perturbations. Unless one makes the very ad-hoc assumption that the duration of inflation is limited to 
the one necessary to produce the CMB modes that we observe at the largest scales (that is, $N_{\rm tot} = N_{\rm CMB} \simeq 60$), also perturbations of wavelength larger than our horizon were produced during inflation with a scale invariant spectrum. Each of these infrared (IR) modes is observed  as a classical and homogeneous vector field on our sky \cite{Bartolo:2012sd}. We can actually observe only the sum of such modes, and this precisely constitutes the homogeneous vector vev ${\bf E}^{\rm vev}$ that we have studied in this work.

If we could observe many disconnected patches of the universe that have a size comparable with our Hubble horizon, we would observe many realizations of the IR gauge modes generated during the first $N_{\rm tot} - N_{\rm CMB}$ e-folds of inflation, and we would observe a different sum  ${\bf E}^{\rm vev}$ of these modes in each Hubble patch. Under the hypothesis of Gaussianity of the gauge perturbations (even a ${\rm O } \left( 1 \right)$ departure from Gaussianity would not change the order of magnitude estimate performed here), the values of these sum would have a Gaussian distribution  of zero mean and variance 
\begin{eqnarray} 
\vert \gamma \vert \gg \frac{1}{2} \;:\;\;\;\;   
\Braket{{\bf E}^2 \left( {\bf x} \right)} 
&=& \frac{1}{2 \, \pi^2} \int_{\rm IR} d k \, k^2 \, \left\vert \delta E_+(k) \right\vert^2 = \frac{9 H^4}{64 \pi^3} \, \frac{{\rm e}^{4 \pi \vert \gamma \vert}}{\vert \gamma \vert^3} \, \int_{\rm IR} \frac{d k}{k} \nonumber\\ 
&=&  \frac{9 H^4}{64 \pi^3} \, \frac{{\rm e}^{4 \pi \vert \gamma \vert}}{\vert \gamma \vert^3} \left( N_{\rm tot} - N_{\rm CMB} \right) \;. 
\end{eqnarray}

We observe only one Hubble patch, and hence only one of such realizations. The variance we have just computed gives the typical value of ${\bf E}_{\rm vev}^2$  obtain in any realization. We therefore obtain 
\begin{equation} 
\vert \gamma \vert \gg \frac{1}{2}  \;:\;\;\;\; 
 \left.  \frac{\rho_E^{\rm vev}}{\rho_\phi} \right\vert_{\rm typical} = \frac{3}{128 \pi^3} \, \left( \frac{H}{M_p} \right)^2  \, \frac{{\rm e}^{4 \pi \vert \gamma \vert}}{\vert \gamma \vert^3} \left( N_{\rm tot} - N_{\rm CMB} \right) \;. 
\label{eq:rat-typical} 
\end{equation} 

The corresponding values of $g_*$ and $c_L$ are  
\begin{eqnarray}
\vert \gamma \vert \gg \frac{1}{2}  \;:\;\;\;\; 
\left. g_* \right|_{\rm typical} &\simeq& 
- 1.3 \times 10^3 \, 
\left( \frac{H}{M_p} \right)^2 \, 
\frac{{\rm e}^{8 \pi |\gamma|}}{|\gamma|^{6}} \, 
\frac{0.01}{\epsilon}
  \, 
\left(\frac{N_{\rm CMB}}{60}\right)^2 \,
\frac{N_{\rm tot} - N_{\rm CMB}}{10}
 ~, \nonumber \\ 
\left. c_0 \right|_{\rm typical} &\simeq& 
4.1 \times 10^3 \, 
 \left( \frac{H}{M_p} \right)^2 \, 
\frac{{\rm e}^{12 \pi |\gamma|}}{|\gamma|^{9}} \,
\frac{0.01}{\epsilon}
  \, 
\left(\frac{N_{\rm CMB}}{60}\right)^3 \,
\frac{N_{\rm tot} - N_{\rm CMB}}{10} ~,
\nonumber \\ 
\left. c_1 \right|_{\rm typical} &=& - \frac{3}{2} \left. c_0 \right|_{\rm typical} \;\;,\;\; 
\left. c_2 \right|_{\rm typical} = \frac{1}{2} \left. c_0 \right|_{\rm typical} \;.  
\end{eqnarray}

Already requiring that the energy in the vector field is subdominant poses a significant limit. For instance, for $|\gamma| = 5.5$, and for only $10$ e-folds of inflation more than the last $N_{\rm CMB}$ ones (this is a conservative assumption, as typical models of inflation give a much longer duration), one finds a subdominant vector field only if $H \la  10^{-13} \, M_p (= 3 \times 10^5 \, {\rm GeV})$. The limit obtained in this work is much stronger. Combining the results \eqref{eq:bound2} and \eqref{eq:rat-typical}, we find the bound 
\begin{equation}
\vert \gamma \vert \gg \frac{1}{2}  \;:\;\;\;\;  \frac{H}{M_p} \la 0.06 \: \frac{\vert \gamma \vert^{9/2}}{{\rm e}^{6 \pi \vert \gamma \vert}} \: \sqrt{\frac{\epsilon}{0.01} \left( \frac{60}{N_{\rm CMB}} \right)^3 \, \frac{10}{N_{\rm tot} - N_{\rm CMB}}} \;, 
\label{eq:final-large-ga}
\end{equation} 
which, for $|\gamma| = 5.5$, gives the bound $\frac{H}{M_p} \la 10^{-43} $. In principle, one can assume values of $\frac{H}{M_p}$ extremely small during inflation, although this is clearly a challenge for model building. The absolute minimum is obtained by imposing an instantaneous thermalization of the inflaton decay products, with the minimum reheating temperature $T_{\rm rh,min} \simeq 4 \, {\rm MeV}$ required for BBN \cite{Hannestad:2004px}. This gives $\frac{H}{M_p} \ga 10^{-42}$. We thus see that values $\vert \gamma \vert \ga 5.5$ are ruled out. 

For comparison, in the $\gamma = 0$ case one finds  \cite{Bartolo:2012sd} 
\begin{eqnarray}
\gamma  = 0   \;:\;\;\;\;  
\left.  \frac{\rho_E^{\rm vev}}{\rho_\phi} \right\vert_{\rm typical} 
&=& \frac{3}{4 \pi^2} \, \frac{H^2}{M_p^2}  \,  \left( N_{\rm tot} - N_{\rm CMB} \right) \;, \nonumber\\ 
\Rightarrow \;\; \frac{H}{M_p} &\la& 4 \times 10^{-5} \: \sqrt{\frac{\epsilon}{0.01} \left( \frac{60}{N_{\rm CMB}} \right)^2 \, \frac{10}{N_{\rm tot} - N_{\rm CMB}}} \;, \label{eq:final-ga0}
\end{eqnarray}
so that a much greater value of $H$ during inflation can be tolerated in this case.

\begin{figure}[t]
  \begin{tabular}{c}
    \begin{minipage}{1.0\hsize}
  \begin{center}
    \includegraphics[width = 1.0\textwidth]{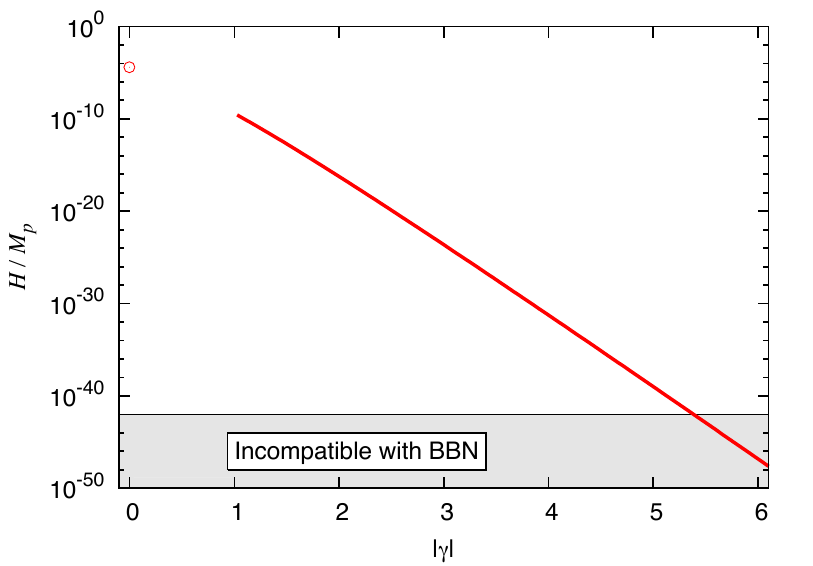}
  \end{center}
\end{minipage}
\end{tabular}
\caption{Upper bounds on $H$ reported in eqs.~\eqref{eq:final-large-ga} and \eqref{eq:final-ga0},  with $N_{\rm CMB} = 60$, $\epsilon = 0.01$ and $N{\rm tot} - N_{\rm CMB} = 10$. One can see that values $ \vert \gamma \vert \gtrsim 5.5$ are ruled out as they would require a reheating temperature smaller than that required for successful BBN.} 
\label{fig:upperbound_gamma_H}
\end{figure}

The bounds \eqref{eq:final-large-ga} and \eqref{eq:final-ga0} are shown in figure \ref{fig:upperbound_gamma_H} for the specific choice of the inflationary parameters $\epsilon = 0.01 ,\, N_{\rm CMB} = 60 ,\, N_{\rm tot} = 70 \,$. Strictly speaking, the result \eqref{eq:final-large-ga} is valid only at $\vert \gamma \vert \gg \frac{1}{2}$ (that is, $\vert \xi \vert \gg 1$).  However, the figure shows that a naive extrapolation of the line from the $\vert \gamma \vert \gg 1$ regime to the  $\vert \gamma \vert \lesssim 1$ one is qualitatively consistent with the result at $\gamma =0$.

\section{Discussion}

In ref.~\cite{Bartolo:2014hwa} we computed the non-diagonal correlators between multipoles of the CMB temperature anisotropy and polarization induced by the simultaneous breaking of parity and rotational invariance during inflation. A very simple way to realize such a breaking is to couple the  inflaton $\phi$ with some gauge field with non-vanishing vev through a pseudo-scalar interaction  $\phi F {\tilde F}$. A vector field with a standard kinetic term is rapidly diluted away by the expansion of the universe. To cope with this, in the present work we considered  the recent modification by Caprini and Sorbo  \cite{Caprini:2014mja} of the  Ratra mechanism \cite{Ratra:1991bn} for the generation of a primordial magnetic field during inflation. This model  is characterized by the interaction $I^2 (\phi) ( -\frac{1}{4} F^2 + \frac{\gamma}{4} F {\tilde F} )$, which explicitly breaks parity. For the suitable time dependence $I ( \phi (t) ) \propto a^{2} (t)$, where $a$ is the scale factor, the model produces scale invariant magnetic perturbations, which is at the core of the mechanisms of ref.~\cite{Ratra:1991bn}.  For $I \propto a^{-2}$ the model produces scale invariant electric perturbations. This was first used in the $\gamma =0$ case to sustain anisotropic inflation \cite{Watanabe:2009ct}, exploiting an ``electromagnetic duality'' in the model  for $I \rightarrow \frac{1}{I} $ \cite{Giovannini:2009xa}. 

We restricted our computations to the  $I \propto a^{-2}$ case; however, due to the duality, we expect that our results can be readily extended to  $I \propto a^{2}$. We also assumed that the field $\phi$ is the inflaton (this is not required in the magnetogenesis applications \cite{Ratra:1991bn,Caprini:2014mja}). In presence of the vector vev, the interaction between the gauge and the inflaton perturbations leads to anisotropic correlators of the primordial perturbations. In particular, among these correlators, we have shown for the first time that it is possible to generate a primordial non-Gaussian signature \emph{during} inflation proportional to $c_1$ in the parametrization~\eqref{eq:BinLegendre}, which is due to broken parity and rotational invariance.

The fact that the anisotropic signatures predicted by this model are not observed in the data results in a  limit on the vev of the vector field, and on the associated energy density $\rho_E^{\rm vev} \equiv \frac{{\bf E}_{\rm vev}^2}{2}$. The limit is already strong at $\gamma =0$, where standard values of the slow roll parameter $\epsilon = 0.01$ and of the number of observable e-folds of inflation $N_{\rm CMB} = 60$ lead to $ \rho_E^{\rm vev} / \rho_\phi \la 10^{-9}$. Such a strong limit is due to the fact that the vector modes act as isocurvature modes that continue to source the inflaton adiabatic perturbations in the super-horizon regime, inducing a $N_{\rm CMB}^2$ ($N_{\rm CMB}^3$) enhancement of the anisotropic two (three) point correlators, and to the fact that the direct inflaton-vector field coupling is stronger than the gravitational one \cite{Bartolo:2012sd}. The limit becomes much more stringent with growing $\vert \gamma \vert$, see figure \ref{fig:95CL_gamma_Ephi}, as the amount of gauge quanta produced by the moving inflaton grows exponentially with $\vert \gamma \vert$ at large $\vert \gamma \vert$. 

The relevant question to ask is whether such stringent limits are compatible with the vev that should naturally be expected in the model. We have considered a mechanism that, by construction, gives scale invariant perturbations for the gauge field. 
In typical inflationary models, the primordial perturbations of size comparable to our horizon  were produced $N_{\rm CMB} \simeq 60$ e-folds before the end of inflation. This does not mean that the scale invariance induced in the mechanism should stop at such wavelengths, but it is natural to assume that also modes produced earlier were scale invariant.~\footnote{One can make ad-hoc assumptions to prevent this: for instance one can postulate that the duration of inflation is only limited to $N_{\rm CMB}$ e-folds, or that the scaling  $I \propto a^{-2}$ did not hold at $N > N_{\rm CMB}$. We do not make such ad-hoc assumptions in discussing what should be naturally expected from this mechanism.} We denote such modes as IR modes. Each of these modes is observed in our sky as classical and homogeneous vector field on our sky \cite{Bartolo:2012sd}. We can actually observe only the sum of such modes, and this precisely constitutes the homogeneous vector vev ${\bf E}^{\rm vev}$ from the IR modes (that eventually sum up to a vev of the vector field from the classical equations of motion) that we have studied in this work.

Let us denote with $N_{\rm tot}$ the total number of e-folds of inflation, and assume that ${\bf E}^{\rm vev} = 0$ at the onset of inflation (if a sizable vev is already present at the start of inflation, we simply expect an even stronger effect, and even stronger bounds). The IR modes that are seen as homogeneous from the point of view of the CMB modes are those produced in the first $ N_{\rm tot} - N_{\rm CMB}$ number of e-folds of inflation. The value of the same on our Hubble patch is obtained as a stochastic addition, obeying the typical random walk relation  $\langle {\bf E}_{\rm vev}^2  \rangle \vert_{\rm expected} \propto N_{\rm tot} - N_{\rm CMB}$ (this is completely analogous \cite{Bartolo:2012sd} to how condensates of light scalar field develop during inflation). This was studied in ref.~\cite{Bartolo:2012sd} for the $\gamma = 0$ case. In the present work, we extended this study to non-vanishing and large $\gamma$, where, due to the significant increase of the amplitude of the produced gauge quanta, the limits imposed by this effect become much more stringent, see figure \ref{fig:upperbound_gamma_H}.  For instance, by simply allowing inflation to last $10$ more e-folds than the amount necessary to produce the CMB modes (this is a very conservative assumption), we could rule out $\vert \gamma \vert \ga 5.5$. 

Our results hold under the assumption that the function $I (\phi)$ produces a constant electric field. This can be achieved for  $I (\phi) \propto a^n$, with $n=-2$.  Using an ``electromagnetic duality'' of the model, this result can be immediately translated to the $I \propto a^2$ case, although for $n>0$ the mechanism is plagued by a strong coupling problem during inflation  \cite{Demozzi:2009fu} (see the discussion after eq.~\eqref{eq:n-def}). On the other hand, the $n \neq -2$ case remains to be studied. Moreover, it would be interesting to study how the limits obtained here weaken in the case in which the field $\phi$ is not the inflaton. We hope to come back to these issues in a separate work.

\acknowledgments
MS was supported in part by a Grant-in-Aid for JSPS Research under Grant Nos.~25-573 and 27-10917, and in part by World Premier International Research Center Initiative (WPI Initiative), MEXT, Japan. This work was supported in part by the ASI/INAF Agreement I/072/09/0 for the Planck LFI Activity of Phase E2. The work of MP was supported in part by DOE grant de-sc0011842 at the University of Minnesota.


\bibliography{paper}
\end{document}